\documentstyle[12pt]{article}
\begin{document}
\def\fnote#1#2{\begingroup\def\thefootnote{#1}\footnote{#2}
\endgroup}

\hfill UTTG-01-09

\hfill TCC-013-09

\begin{center}
\Large
Living With Infinities\\
\normalsize 
\vspace{10pt}
Steven Weinberg\fnote{*}{Electronic address:
weinberg@physics.utexas.edu}\\
{\em Theory Group, Department of Physics, and Texas Cosmology Center,\\ University of
Texas,
Austin, TX, 78712}

\vspace{30pt}

\noindent
{\bf Abstract}
\end{center}
\noindent
This is the written version of a talk given in memory of Gunnar K\"{a}ll\'{e}n, at the Departments of Theoretical Physics, Physics, and Astronomy of Lund University on February 13, 2009.  It will be published in a  collection of the papers of Gunnar K\"{a}ll\'{e}n, edited by C. Jarlskog and A. C. T. Wu.  I discuss some of K\"{a}ll\'{e}n's  work,  especially regarding the problem of infinities in quantum field theory, and recount my own interactions with him.  In addition, I describe for non-specialists the current status of the problem, and  present my personal view on how it may be resolved in the future.

\vfill

\pagebreak

\setcounter{footnote}{0}

I owe a great debt of gratitude to Gunnar K\"{a}ll\'{e}n.  In the summer of 1954, having just finished my undergraduate studies at Cornell, I arrived at the Bohr Institute in Copenhagen, where K\"{a}ll\'{e}n was a member of the Theoretical Study Division of CERN, which had not yet moved to Geneva.  Richard Dalitz had advised me to go to Copenhagen partly because of the presence there of CERN.  But my real reason for coming to Copenhagen with my wife was that we had just married, and thought that we could have a romantic year abroad before we returned to the U.S. for me to enter graduate school.  I brought with me a bag of physics books to read, but I did not imagine that I could start original research.  You see, I had the idea that before I started research on any topic, I first had to know everything that had been done in that area, and I knew that I was far from knowing everything about anything.  

It wasn't long before people at the Institute let me know that everyone there was expected to be working on some sort of research.  David Frisch, a visiting American nuclear physicist, kindly suggested that I do something on nuclear alpha decay, but nothing came of it.  

Early in 1955  I heard that a young theorist named K\"{a}ll\'{e}n was doing interesting things in  quantum field theory, so I knocked on his office door, and asked him to suggest a research problem.  As it happened, K\"{a}ll\'{e}n did have a problem to suggest.  A year earlier,  Tsung-Dao Lee at Columbia had invented a clever field-theoretic model that could be solved exactly.\footnote{T.D. Lee, Phys. Rev. 95, 1329 (1954).}   The model had some peculiarities, which I'll come back to.  These problems did not at first  seem fatal to Lee, but  K\"{a}ll\'{e}n  joined with the great Wolfgang Pauli to show that scattering processes in the Lee model violate the principle of unitarity --- that is, the sum of the probabilities for all the things that can happen when two particles collide did not always add up to 100\%.\footnote{G. K\"{a}ll\'{e}n and W. Pauli, Dan.  Mat. Fys. Medd. 30, no. 7 (1955).}  Now K\"{a}ll\'{e}n wanted me to see if there were other things wrong with the Lee model.  

With a great deal of patient help from K\"{a}ll\'{e}n, I was able to show that there were states in the Lee model whose energies were complex --- that is, not ordinary real numbers.  I finished the work on the Danish freighter that took my wife and me back to the U.S., and soon after I started graduate school at Princeton I had published the work as my first research paper.\footnote{S. Weinberg, Phys. Rev. 102, 285 (1956).}  This was a pretty unimportant paper (I recently checked, and found that it has been cited just nine times in 53 years), but it was a big thing for me --- I started to feel like a physicist, not a student.  

Incidentally, K\"{a}ll\'{e}n's kindness to me went beyond starting me in research.  He and his wife had my wife and me to their house for dinner, and going to the bathroom there, I learned something about K\"{a}ll\'{e}n that probably most of you don't know --- he had hand towels embroidered with the Dirac equation.  Mrs. K\"{a}ll\'{e}n told me  that they were a present from Pauli. K\"{a}ll\'{e}n also introduced me to Pauli, but I didn't get any towels.

Even though I had benefited so much from K\"{a}ll\'{e}n's suggestion of a research problem, I felt that there was something odd about it.   Lee was then not a well-known theorist --- his great work with Yang on parity violation and weak interactions was a few years in the future.  Also, the Lee model was not intended to be a serious model of real particles.  So why did K\"{a}ll\'{e}n  take the trouble to shoot it down, even to the extent of enlisting the collaboration of his friend Pauli?  The explanation, which I understood only much later,  has to do with a long-standing controversy about the future of quantum field theory, in which K\"{a}ll\'{e}n was playing an important part.

The controversy concerned the significance of infinities in quantum field theory.  The problem of infinities was anticipated in the first papers on quantum field theory by Heisenberg and Pauli,\footnote{W. Heisenberg and W. Pauli, Z. f. Physik 56, 1 (1929); 59, 168 (1930).} and then in 1930 infinite energy shifts were found in calculations of the effects of emitting and reabsorbing photons by free or bound electrons, by Waller\footnote{I. Waller, Z. f. Physik 59, 168 (1930); 61, 721, 837 (1930); 62, 673 (1930)} and Oppenheimer.\footnote{J. R. Oppenheimer, Phys. Rev. 35, 461 (1930).}  In both cases you have to integrate over the momenta of the photons, and the integrals diverge.  During the 1930s it was widely believed that these infinities signified a breakdown of quantum electrodynamics at energies above a few MeV.  This changed after the war, when new techniques of calculation were developed that manifestly preserved the principles of special relativity at every step, and it was recognized that the infinities could be absorbed into a redefinition, called a {\em renormalization}, of physical constants like the charge and mass of the electron.\footnote{See articles by Bethe, Dyson, Feynman, Kramers, Lamb \& Retherford, Schwinger,  Tomonaga, and Weisskopf reprinted in {\em Quantum Electrodynamics}, ed. J. Schwinger (Dover Publications, Inc., New York, 1958).}  Dyson was able to show (with some technicalities cleared up later by Salam\footnote{A. Salam, Phys. Rev. 82, 217 (1951).} and me\footnote{S. Weinberg, Phys. Rev. 118, 838 (1959).}) that in quantum electrodynamics and a limited class of other theories, the renormalization of a finite number of physical parameters would actually remove infinities in every order of perturbation theory --- that is, in every term when we write any physical observable as an expansion in powers of the charge of the electron, or powers of similar parameters in other theories.  Theories in which infinities are removed in this way are known as {\em renormalizable.}  They can be recognized by the property that in renormalizable theories, in natural units in which Planck's constant and the speed of light are unity,  all of the  constants multiplying terms in the Lagrangian are just pure numbers, like the charge of the electron, or have the units of positive powers of energy, like particle masses, but not negative powers of energy.\footnote{The units of these constants of course depend on the units we assign to the field operators.  In using this criterion for renormalizability, it is essential to use units for any field operator related to the asymptotic behaviour of its propagator; if the propagator goes like $k^n$ for large four-momentum $k$, then the field must be assigned the units of energy to the power $n/2+2$.  In particular, because of $k^\mu k^\nu/(k^2+m^2)$ terms in the propagator of a massive vector field, for these purposes the field must be given the unconventional units of energy to the power $+2$, and any interaction of the field would be non-renormalizable, unless the field is coupled only to conserved currents for which the terms in the propagator proportional to $k^\mu k^\nu$  may be dropped.}

The great success of calculations in quantum electrodynamics using the renormalization idea  generated a new enthusiasm for quantum electrodynamics.  After this change of mood, probably most theorists simply didn't worry about having to deal  with infinite renormalizations.  Some theorists thought that these infinities were just a consequence of having expanded in powers of the electric charge of the electron, and that not only observables but even quantities like the ``bare'' electron charge  (the charge appearing in the field equations of quantum electrodynamics) would be found to be finite when we learned how to calculate without perturbation theory.  But at least two leading theorists had their doubts about this, and thought that the appearance of infinite renormalizations in perturbation theory was a symptom of a deeper problem, a problem not with  perturbation theory but with quantum field theory itself.  They were Lev Landau, and Gunnar K\"{a}ll\'{e}n.  

K\"{a}ll\'{e}n's first step in exploring this problem was in an important 1952 paper,\footnote{G. K\"{a}ll\'{e}n, Helv. Phys. Acta 25, 417 (1952).} in which  he showed  how to define quantities like the bare charge of the electron without the use of perturbation theory.    To avoid the complications that arise from the vector nature of the electromagnetic field, I'll describe the essential points here using the easier example of a real scalar field $\varphi(x)$, studied a little later by Lehmann.\footnote{H. Lehman, Nuovo Cimento XI, 342 (1954).}   The quantity $-i\Delta'(p)$ known as the propagator, that in perturbation theory would be given by the sum of all Feynman diagrams with two external lines, carrying four-momenta $p^\mu$ and   $-p^\mu$,   can be defined without the use of perturbation theory by
\begin{equation}
\Big\langle 0 \Big| T\{\varphi(x)\;,\;\varphi(0)\Big\}\Big| 0\Big\rangle = -i\int \frac{d^4p}{(2\pi)^4}\Delta'(p) e^{i p\cdot x}\;,
\end{equation}
where $|0\rangle$ is the physical vacuum state, and $T$ denotes a time-ordered product, with $\varphi(x)$ to the left or right of $\varphi(0)$ according as the time $x^0$ is positive or negative.  By inserting a complete set of states between the fields in the time-ordered product, one finds  what has come to be called the K\"{a}ll\'{e}n--Lehmann representation
\begin{equation}
\Delta'(p)=\frac{|N|^2}{p^2+m^2}+\int \frac{\sigma(\mu)\;d\mu}{p^2+\mu^2}\;,
\end{equation}
where $\sigma(\mu^2)\geq 0$ is given by a sum over multiparticle states with total energy-momentum vector $P^\lambda$ satisfying $-P^2=\mu^2$, and  $N$ is defined by the matrix element of $\varphi(x)$ between the vacuum and a one-particle state of physical mass $m$ and three-momentum ${\bf k}$:
\begin{equation}
\Big\langle 0 \Big|\varphi(x)\Big|{\bf k}\Big\rangle=\frac{N e^{ik\cdot x}}{(2\pi)^{3/2}\sqrt{2k^0}}\;,
\end{equation}
with $k^0\equiv  \sqrt{{\bf k}^2+m^2}$.  
If $\varphi(x)$ is the ``unrenormalized'' field that appears in the quadratic part of the Lagrangian without any extra factors, then it satisfies the canonical commutation relation
\begin{equation}
\Big[\dot{\varphi}({\bf x},t)\,,\,\varphi({\bf y},t)\Big]=-i\delta^3({\bf x}-{\bf y})\;.
\end{equation}
By taking the time derivative of Eq.~(1) and then setting the time $x^0$ equal to zero and using the commutation relation (4), one obtains the sum rule
\begin{equation}
1=|N|^2+\int \sigma(\mu)\,d\mu \;.                                
\end{equation}

One immediate consequence is that, since $|N|^2$ is necessarily positive, Eq.~(5) gives an upper limit on the coupling of the field $\varphi$ to multiparticle states
\begin{equation}
\int \sigma(\mu)\,d\mu\leq 1 \;.
\end{equation}
I'll mention in passing that this  upper limit is reached in the case $N=0$, which only applies if $\varphi(x)$ does not appear in the Lagrangian at all --- that is, if the particle in question is not elementary.  Thus, in a sense, composite particles are coupled to their constituents more strongly than any possible elementary particle.

This kind of sum rule has proved very valuable in theoretical physics.  For instance, 
if instead of a pair of scalar fields in Eq.~(1) we consider  pairs of  conserved symmetry currents, then by using methods similar to  K\"{a}ll\'{e}n's, one gets what are called a spectral function sum rules,\footnote{S. Weinberg, Phys. Rev. Lett. 18, 507 (1967).} which have had useful applications, for instance in calculating the decays of vector mesons into electron--positron pairs.

What chiefly concerned K\"{a}ll\'{e}n was the application of these methods to quantum electrodynamics.  In his 1952 paper, K\"{a}ll\'{e}n derived a sum rule like (5) for the electromagnetic field, with $Z_3\equiv |N_\gamma|^2$ in place of $|N |^2$, where $N_\gamma$ is the renormalization constant for the electromagnetic field.  As in the scalar field theory, this sum rule (and the definition of $Z_3$ as an absolute value squared) shows that
\begin{equation}
0\leq Z_3<1\;.
\end{equation}
This is especially important in electrodynamics, because $Z_3$ appears in the relation between the bare electronic charge $e_B$ that appears in the field equations, and the physical charge $e$ of the electron:
\begin{equation}
e^2=Z_3\,e_B^2\;.
\end{equation}
The fact that $e^2$ is less than $e^2_B$ has a well-known interpretation: it is due to the shielding of the bare charge by virtual positrons, which are pulled out of the vacuum along with virtual electrons, and unlike the virtual electrons are attracted to the real electron whose charge is being measured.
  
Now, in  lowest order perturbation theory, we have
\begin{equation}
Z_3=1-\frac{e^2}{6\pi^2}\ln\left(\frac{\Lambda}{m_e}\right)\;,
\end{equation}
where $\Lambda$ is an ultraviolet cut-off, put in as a limit on the energies of the virtual photons.  This is all very well if we take $\Lambda$ as a reasonable multiple of the electron mass $m_e$, but if the cut-off is taken greater than $m_e\exp(6\pi^2/e^2)\approx 10^{280}m_e$ (which is more than the total mass of the observable universe) then we are in trouble: In this case Eq.~(9) gives $Z_3$  negative, contradicting the inequality (7).  As Landau pointed out,\footnote{L. Landau, in {\em Niels Bohr and the Development of Physics} (Pergamon Press, New York, 1955): p. 52.} this ridiculously large energy becomes much smaller if we take into account the fact that there are several species of charged elementary particles; for instance, if there are $\nu$ species of spin one-half particles with the same charge as the electron, then the factor $10^{280}$ is replaced with $10^{280/\nu}$.  So if $\nu$ is, say,  10 or 20, the problem with the sign of $Z_3$ would set in at energies much closer to those with which we usually have to deal.  But this is just lowest order perturbation theory --- to see if there is really any problem, it is necessary to go beyond perturbation theory.

To explore this issue, K\"{a}ll\'{e}n set out to see if the integral appearing in $1-Z_3$, and not just its expansion  in powers of $e^2$, actually diverges in the absence of a cut-off.  Of course, he could not evaluate the integral exactly, but since every kind of multiparticle state makes a positive contribution to the integrand, he could  concentrate on the contribution of the simplest states, consisting of just an electron and a positron --- if the integral of this contribution diverges, then the whole integral diverges.  In evaluating this contribution, he had to assume that all renormalizations including the renormalization of the electron mass and field were finite.  With this assumption, and some tricky interchanges of integrations, he found that the integral for $1-Z_3$ does diverge.  In this way, he reached his famous conclusion that at least one of the renormalization constants in quantum electrodynamics has to be infinite.\footnote{G. K\"{a}ll\'{e}n, Dan. Mat. Fys. Medd. 27, no. 12 (1953).}  

Not everyone was convinced.  To quote the K\"{a}ll\'{e}n memorial statement of Paul Urban in 1969,\footnote{P. Urban, Acta Physica Austriaca, Suppl. 6 (1969).} ``Indeed, other authors are in doubt about his famous proof that at least one of the renormalization constants has to be infinite, but so far no definite answer to this question has been found.'' It should be noted that at the end of his 1953 paper, K\"{a}ll\'{e}n had explicitly disavowed any claim to mathematical rigor.  As far as I know, this issue has never been settled.  Of course, the important question was not whether some of the renormalization constants are infinite for infinite cut-off, but whether something happens at very high energies, such as $10^{280}m_e$, to prevent the cut-off in  quantum electrodynamics from being taken to infinity.    I don't know if K\"{a}ll\'{e}n ever expressed an opinion about it, but I suspect that he thought that quantum electrodynamics does break down at very high energies, and that he wanted to be the one who proved it.

Which brings me back to the Lee model.  This is a model with two heavy particles, $V$ and $N$, and a lighter particle $\theta$, all with zero spin.  The only interactions in the theory are ones in which $V$ converts to $N+\theta$, or vice versa.  No antiparticles are included, and the recoil energies of the $V$ and $N$ are neglected, so the model is non-relativistic, though the energy $\omega$ of a $\theta$ of momentum ${\bf p}$ is given by the relativistic formula $\omega=\sqrt{{\bf p}^2+m_\theta^2}$.  The model is exactly soluble in sectors with just one or two particles.  For instance, to find the complete amplitude for $V\rightarrow N+\theta$, one  can sum the graphs for 
$$ V\rightarrow N+\theta\rightarrow V\rightarrow N+\theta\rightarrow V\rightarrow \cdots\rightarrow N+\theta\;,$$
which is just a geometric series.  One finds that, if the  physical and bare $V$-particle states are normalized so that
\begin{equation}
\langle V, {\rm phys} | V,{\rm phys} \rangle=\langle V, {\rm bare} | V, {\rm bare} \rangle=1\;,
\end{equation}
then  we have an exact sum rule resembling (5): 
\begin{equation}
1=|N|^2+\frac{|g|^2}{4\pi^2}\int_0^\Lambda \frac{k^2\,dk}{\omega^3}\;,
\end{equation}
where 
\begin{equation}
N\equiv \langle V, {\rm bare} | V,{\rm phys} \rangle
\end{equation}
Here $\Lambda$ is again an ultraviolet cut-off, and $g$ is the renormalized coupling for this vertex, 
 related to the bare coupling $g_B$ by the exact formula $g=Ng_B$.
 For $\Lambda\gg m_\theta$, the integral in Eq.~(11) grows as $\ln\Lambda$, so if $g\neq 0$ then $\Lambda$ cannot be arbitrarily large without violating 
the condition  that $|N|^2\geq 0$.  This is just like the problem encountered in lowest-order quantum electrodynamics, except that here there is no use of perturbation theory, and hence no hope that the difficulty will go away when perturbation theory is dispensed with.

Despite this difficulty, Lee found that his model with $\Lambda\rightarrow \infty$  gave sensible results for some simple problems, like the calculation of the energy of the $V$ particle.  In their 1955 paper, K\"{a}ll\'{e}n and Pauli confronted the difficulty that $|N|^2$ then comes out negative, and recognized that for very large $\Lambda$  this was necessarily a theory with an indefinite metric --- that is, it is necessary to take all states with odd numbers of bare $V$ particles with negative norm, while all other states with definite numbers of bare particles have positive norm.  In particular, in place of (10), we must take $\langle V, {\rm bare} | V, {\rm bare} \rangle=-1$, while calculations show that the physical $V$ state has positive norm, so that we can still normalize it so that $
\langle V, {\rm phys} | V,{\rm phys} \rangle=+1$.  (There is also another discrete energy eigenstate formed as a superposition of bare $V$ and $N+\theta$ states, that has  negative norm.)  Then in place of (11), we have
\begin{equation}
1=-|N|^2+\frac{g^2}{4\pi^2}\int_0^\Lambda \frac{k^2\,dk}{\omega^3}\;,
\end{equation}
which gives no problem for large $\Lambda$.  The device of an indefinite metric had already been introduced   by Dirac,\footnote{P. A. M. Dirac, Proc. Roy. Soc. A180, 1 (1942).} for reasons having nothing to do with infinities (Dirac was trying to find a physical interpretation of the negative energy solutions of the relativistic wave equations for bosons), and  Pauli\footnote{W. Pauli, Rev. Mod. Phys. 15, 175 (1943).} had noticed that if we can introduce suitable negative signs into sums over states, it should be possible to avoid infinities altogether.  I think that what  K\"{a}ll\'{e}n and Pauli in 1955 disliked about the indefinite metric was not that it solved the problem of infinities, but that it did so too easily, without having to worry about what really happens at very high energies and short distances, and this is why they took the trouble to show that it did lead to unphysical results in the Lee model.

Experience has justified K\"{a}ll\'{e}n and Pauli's distrust of the indefinite metric.  This device continues to appear in theoretical physics, but only where there is some symmetry principle that cancels the negative probability for producing states with negative norm by the positive probability for producing other unphysical states, so that the total probability of producing physical states still adds up to 100\%.    Thus, in the Lorentz-invariant quantization of the electromagnetic field by Gupta and Bleuler,\footnote{S. N. Gupta, Proc. Phys. Soc. 58, 681 (1950); K. Bleuler, Helv. Phys. Acta 28, 567 (1950).} the state of a timelike photon has negative norm, but gauge invariance insures that the negative probability for the production of these unphysical photons with timelike polarization is canceled by the positive probability for the production of  other unphysical photons, with longitudinal polarization.    A similar cancelation occurs in the Lorentz invariant quantization of string theories, where the symmetry is conformal symmetry on the two-dimensional worldsheet of the string.  But it seems that without any such symmetry, as in the Lee model, the indefinite metric does not work.\footnote{It has been argued that the ${\sf PT}$ symmetry of the Lee model allows the definition of a scalar product for which the theory is unitary; see C. M. Bender, S. F. Brandt, J-H Chen, and Q. Wang, Phys. Rev. D 71, 025014 (2005); C. M. Bender and P. D. Mannheim, Phys. Rev. D 78, 025022 (2008).}  

I should say a word about where we stand today regarding the survival of quantum electrodynamics and other field theories in the limit of very high cut-off.  The appropriate formalism for addressing this question is the renormalization group formalism presented by Wilson\footnote{K. G. Wilson, Phys. Rev. B4, 3174, 3184 (1971); Rev. Mod. Phys. 47, 773 (1975).} in 1971.  When  we calculate the logarithmic derivative of the bare electron charge $e_{B\Lambda}$ with respect to the cut-off $\Lambda$ at a fixed renormalized charge, then the result for $\Lambda \gg m_e$ can only depend on $e_{B\Lambda}$, since there is no relevant quantity with the units of energy with which $\Lambda$ can be compared.  That is, $e_{B\Lambda}$ satisfies a differential equation of the form
\begin{equation}
\Lambda \frac{d e_{B\Lambda}}{d\Lambda}=\beta(e_{B\Lambda})\;.
\end{equation}
The whole question then reduces to the behavior of the function $\beta(e)$.  If it is positive and increases fast enough so that $\int^\infty de/\beta(e)$ converges, then the cut-off in quantum electrodynamics cannot be extended to a value greater than a finite energy $E_\infty$, given by
\begin{equation}
E_\infty=\mu\exp\left(\int^\infty_{e_{B\mu}} \frac{de}{\beta(e)}\right)\;,
\end{equation}
with $\mu$ arbitrary.    
On the basis of an approximation in which in each order of perturbation theory one keeps only terms with the maximum number of large logarithms, Landau concluded in ref. 14 that quantum electrodynamics does break down at very high energy.     In effect, he was arguing on the basis of the lowest-order term, $\beta(e)\simeq e^3/12\pi^2$, for which  $\int^\infty de/\beta(e)$  does converge.

No one today knows whether this is the case.    It is equally possible that higher-order effects will make $\beta(e)$ increase more slowly or even decrease for very large $e$, in which case 
$\int^\infty de/\beta(e)$ will diverge and $e_{B\Lambda}$ will just continue to grow smoothly with $\Lambda$.  One might imagine that  $\beta(e)$ could instead drop to zero at some finite value $e_*$, in which case $e_{B\Lambda}$ would approach $e_*$ as $\Lambda\rightarrow \infty$, though there are  arguments against this.\footnote{S. L. Adler, C. G. Callan, D. J. Gross, and R. Jackiw, Phys. Rev. D6, 2982 (1972); M. Baker and K. Johnson, Physica 96A, 120 (1979); P. C. Argyres, M. Ronen, N. Seiberg, and E. Witten, Nucl. Phys. B461, 71 (1996).}  
Lattice calculations (in which spacetime is replaced by a lattice of separate points, providing an ultraviolet cut-off equal to the inverse lattice spacing) indicate that the beta function for  a scalar field theory with interaction $g_B\varphi^4$ increases for large $g_B$ fast enough so that $\int dg_B/\beta(g_B)$ converges and the theory therefore does not have a continuum limit for zero lattice spacing.\footnote{For a discussion and references, see J. Glimm and A. Jaffe, {\em Quantum Physics -- A Functional Integral Point of View}, 2nd ed. (Springer-Verlag, New York, 1987), Sec. 21.6; R. Fernandez, J. Fr\"{o}lich, and A. D. Sokal, {\em Random Walks, Critical Phenomena, amd Triality in Quantum Field Theory} (Springer-Verlag, Berlin, 1992), Chapter 15.}  And in the Lee model without an indefinite metric, Eq.~(11) together with the relation $g_B=g/N$ gives 
$$\beta(g_{B\Lambda})\equiv \Lambda \frac{d g_{B\Lambda}}{d\Lambda}=\frac{g^3_{B\Lambda}}{8\pi^2}$$ 
for $\Lambda\gg m_\theta$, so $\int^\infty dg/\beta(g)$ converges, and as we have seen, the cut-off cannot be taken to infinity.

If limited to quantum electrodynamics, the problem of high energy behavior has become academic,  since electromagnetism merges with the weak interactions at energies above 100 GeV, and we really should be asking about the high energy behavior of the $SU(2)$ and $U(1)$ couplings of the electroweak theory.  Even that is somewhat academic, because  gravitation becomes important at an energy of order $10^{19}$ GeV, well below the energy at which the $SU(2)$ and $U(1)$ couplings would become infinite.  And there is no theory of gravitation that is renormalizable in the Dyson sense --- the Newton constant appearing in General Relativity has the units of an energy to the power $-2$.

K\"{a}ll\'{e}n's concern with the problems of quantum field theory at very high energy did not keep him from appreciating the great success of quantum electrodynamics.  In a contribution to the 1953 Kamerlingh Onnes Conference,\footnote{ G. K\"{a}ll\'{e}n, Physica XIX, 850 (1953.} he remarked that ``there is little doubt that the mathematical framework of quantum electrodynamics contains something which corresponds closely to physical reality.''  He did practical calculations using perturbation theory in quantum electrodynamics, on problems such as the vacuum polarization in fourth order\footnote{G. K\"{a}ll\'{e}n and A. Sabry, Dan. Mat. Fys. Medd. 29, no. 7 (1955).} and the radiative corrections to decay processes.\footnote{G. K\"{a}ll\'{e}n, Nucl. Phys. B 1, 225 (1967).}  He wrote a book about quantum electrodynamics,\footnote{G. K\"{a}ll\'{e}n, {\em Quantum Electrodynamics}, transl.~C. K. Iddings and M. Mizushima (Springer-Verlag, 1972).}   leaving for the very end of the book his concern about the infinite value of renormalization constants.  

K\"{a}ll\'{e}n's interests were not limited to quantum electrodynamics.  In 1954 he showed that the renormalizable meson theory with pseudoscalar coupling could not be used to account for both pion scattering and pion photoproduction, because different values of the pion-nucleon coupling constant are needed in the two cases.\footnote{G. K\"{a}ll\'{e}n, Nuovo Cimento XII, 217 (1954).}  Again, this result relied on lowest-order perturbation theory, so K\"{a}ll\'{e}n acknowledged that it did not conclusively kill this meson theory.  He remarked that ``It would certainly be felt as a great relief by many theoretical physicists --- among them the present author --- if a definite argument against meson theory in its present form or a definite mathematical inconsistency in it could be found. This feeling together with wishful thinking must not tempt us to accept as conclusive evidence an argument that is still somewhat incomplete.'' 

Of course, K\"{a}ll\'{e}n was right in his dislike of this particular meson theory.  A decade or so later the development of chiral Lagrangians showed that low energy pions are in fact well described by a theory with {\em pseudovector} coupling of single pions to nucleons, plus terms with two or more pions interacting with a nucleon at a single vertex, as dictated by a symmetry principle, chiral symmetry.\footnote{For a discussion with references to the original literature, see Sec. 19.5 of S. Weinberg, {\em The Quantum Theory of Fields}, Vol. II (Cambridge Univ. Press, 1996.)}  This theory is not renormalizable in the Dyson sense, but we have learned how to live with that.  It is an effective field theory, which can be used to generate a series expansion for soft pion scattering amplitudes in powers of the pion energy.  The Lagrangian for the theory contains every possible interaction that is allowed by the symmetries of the theory, but the non-renormalizable interactions whose coupling constants are negative powers of some characteristic energy (which is about 1 GeV in this theory) make a small contribution for pion energies that are much less than the characteristic energy.  To any given order in pion energy, all infinities can be absorbed in the renormalization of a finite number of coupling parameters, but we need more and more of these parameters to absorb infinities as we go to higher and higher powers of pion energy.

My own view is that all of the successful field theories of which we are so proud --- electrodynamics, the electroweak theory, quantum chromodynamics, and even General Relativity --- are in truth effective field theories, only with a much larger characteristic energy, something like the Planck energy, $10^{19}$ GeV.  It is somewhat of an accident  that the simplest  versions of electrodynamics, the electroweak theory, and quantum chromodynamics  are renormalizable in the Dyson sense, though it is very important from a practical point of view, because the renormalizable interactions dominate at ordinary accessible energies.  An effect of one of the non-renormalizable terms has recently been detected: An interaction involving two lepton doublets and two scalar field doublets generates neutrino masses when the scalar fields acquire expectation values.\footnote{S. Weinberg, Phys. Rev. Lett. 43, 1566 (1979).}  

None of the renormalizable versions of these theories really describes nature at very high energy, where the non-renormalizable terms in the theory are not suppressed.    
From this  point of view, the fact that General Relativity is not renormalizable in the Dyson sense is no more (or less) of a fundamental problem than the fact that  non-renormalizable terms are present along with the usual renormalizable terms of the Standard Model.  All of these theories lose their predictive power at a sufficiently high energy.  The challenge for the future is to find the final underlying theory, to which the effective field theories of the standard model and General Relativity are low-energy approximations.  

It is possible and perhaps
 likely that the ingredients of the underlying theory are not the quark and lepton and gauge boson fields of the Standard Model, but something quite different, such as strings.  After all, as it has turned out, the ingredients of our modern theory of strong interactions are not the nucleon and pion fields of K\"{a}ll\'{e}n's time, but quark and gluon fields, with an effective field theory of nucleon and pion fields useful only as a low-energy approximation.  

But there is another possibility.   The underlying theory may be an ordinary quantum field theory, including fields for gravitation and the ingredients of the Standard Model.  Of course, it could not be renormalizable in the Dyson sense, so to deal with infinities every possible interaction allowed by symmetry principles would have to be present, just as in effective field theories like the chiral theory of pions and nucleons.  But it need not lose its predictive power at high energies, if the bare coupling constants $g_{n}(\Lambda)$ for an ultraviolet cut-off $\Lambda$ (multiplied by whatever positive or negative powers of $\Lambda$ are needed to make the $g_n$  dimensionless) approach a fixed point $g_{n*}$ as $\Lambda\rightarrow \infty$.\footnote{S. Weinberg, in {\em Understanding the Fundamental Constituents of Matter -- 1976 Erice Lectures}, ed. A. Zichichi (Plenum Press); and in {\em General Relativity}, ed.
S. W. Hawking and W. Israel (Cambridge University Press, 1979) 790.}  This is what happens in quantum chromodynamics, where $g_*=0$, and in that case is known as asymptotic freedom.\footnote{D. J. Gross and F. Wilczek, Phys. Rev. Lett. 40, 1343 (1973); H. D. Politzer, Phys. Rev. Lett. 30, 1346 (1973).}  In theories involving gravitation it is not possible for all the $g_{n*}$ to vanish.  In this more general case where $g_{n*}$ is not necessarily zero, the approach to a fixed point is known as ``asymptotic safety,'' because the theory is safe from the danger that dimensionless couplings like $g_{\rm grav}=G\Lambda^2$ (where $G$ is Newton's constant) might run off to infinity as $\Lambda$ goes to infinity.

  For asymptotic safety to be possible, it is necessary that $\beta_n(g_*)=0$, where $\beta_n(g(\Lambda))\equiv \Lambda\, dg_n(\Lambda)/d\Lambda$.  It is also necessary that the  coupling constants  $g_{n}(\Lambda)$ at any finite cut-off lie on a trajectory in coupling constant space that is attracted rather than repelled by this fixed point.  There are reasons to expect that, even with an infinite number of coupling parameters, the surfaces spanned by such trajectories have finite dimensionality, so such a theory would involve just a finite number of free parameters, just as for ordinary renormalizable theories.  The trouble, of course, is that there is no reason to expect the $g_{n*}$ to be small, so that ordinary perturbation theory can't be relied on for calculations in asymptotically safe theories.  Other techniques such as dimensional continuation,\footnote{S. Weinberg, ref. 31 (1979); H. Kawai, Y. Kitazawa, \& M. Ninomiya, Nucl. Phys. B 404, 684 (1993);   Nucl. Phys. B 467, 313 (1996); T.  Aida \& Y. Kitazawa, Nucl. Phys. B 401, 427 (1997);  M. Niedermaier, Nucl. Phys. B 673, 131 (2003) .}
 $1/N$ expansions,\footnote{L. Smolin, Nucl. Phys. B208, 439 (1982);
 R. Percacci, Phys. Rev. D 73, 041501 (2006).} lattice quantization,\footnote{J. Ambj\o rn, J. Jurkewicz, \& R. Loll, Phys. Rev. Lett. 93, 131301 (2004);  Phys. Rev. Lett. 95, 171301 (2005); Phys. Rev. D72, 064014 (2005);   Phys. Rev. D78, 063544 (2008); and  in {\em Approaches to Quantum Gravity}, ed. D. Or\'{i}ti (Cambridge University Press).}  and the truncated ``exact'' renormalization group equations,\footnote{M. Reuter, Phys. Rev. D 57, 971 (1998); D. Dou \& R. Percacci, Class. Quant. Grav. 15, 3449 (1998); W. Souma, Prog. Theor. Phys. 102, 181 (1999); O. Lauscher \& M. Reuter, Phys. Rev. D 65, 025013 (2001); Class. Quant. Grav. 19. 483 (2002);  M. Reuter \& F. Saueressig, Phys Rev. D 65, 065016  (2002); O. Lauscher \& M. Reuter, Int. J. Mod. Phys. A 17, 993 (2002);  Phys. Rev. D 66, 025026 (2002); M. Reuter and F. Saueressig, Phys Rev. D 66, 125001 (2002); R. Percacci \& D. Perini, Phys. Rev. D 67, 081503 (2002);  Phys. Rev. D 68, 044018 (2003); D. Perini, Nucl. Phys. Proc. Suppl. C 127, 185 (2004); D. F. Litim, Phys. Rev. Lett. {\bf 92}, 201301 (2004); A. Codello \& R. Percacci, Phys. Rev. Lett. 97, 221301 (2006); A. Codello, R. Percacci, \& C. Rahmede, Int. J. Mod. Phys. A23, 143 (2008);  M. Reuter \& F. Saueressig, 0708.1317; P. F. Machado and F. Saueressig, Phys. Rev. D77, 124045 (2008); A. Codello, R. Percacci, \& C. Rahmede, 0805.2909;  A. Codello \& R. Percacci, 0810.0715; D. F. Litim 0810.3675; H. Gies \& M. M. Scherer, 0901.2459; D. Benedetti, P. F. Machado, \& F. Saueressig, 0901.2984, 0902.4630; M. Reuter \& H. Weyer, 0903.2971.} have provided increasing evidence that gravitation may be part of an asymptotically safe theory.
\footnote{For reviews see  M. Niedermaier \& M. Reuther, Living Rev. Relativity 9, 5 (2006); M. Niedermaier, Class. Quant. Grav. 24, R171 (2007); 
 M. Reuter and F. Saueressig, 0708.1317; R. Percacci, in {\em Approaches to Quantum Gravity}, ed. D. Or\'{i}ti (Cambridge University Press).}  So it is just possible that                                                                                                                                                                                                                                                                                                                                                                                                                                                                                                              we may be closer to the final underlying theory than is usually thought.

                                    K\"{a}ll\'{e}n continued his interest in general elementary particle physics, and wrote a book about it, published in 1964.\footnote{G. K\"{a}ll\'{e}n, {\em Elementary Particle Physics} (Addison-Wesley, Reading, MA. 1964).} Arthur Wightman quoted a typical remark about this book: ``That is the book on elementary particles that experimentalists find really helpful.''  But K\"{a}ll\'{e}n's  timing was unlucky -- the development not only of chiral dynamics but also of the electroweak theory were then just a few years in the future, and they were to put many of the problems he worried about in a new perspective.

It was a tragic loss not only to his friends and family but also to all theoretical physics that K\"{a}ll\'{e}n died in an airplane  accident just 40 years ago.  For me, this was specially poignant, because he had been so kind to me in Copenhagen, and yet we had become estranged.  Some time in 1957, just before I finished my graduate work, K\"{a}ll\'{e}n visited Princeton, and left a note in my mail box.  Apparently he had seen a draft of my Ph.~D. thesis, which was about the use of renormalization theory to deal with strong interaction effects in weak decay processes.  His note seemed angry, and said that my work showed all the misconceptions about quantum field theory that were then common.  Well, my thesis was no great accomplishment, but I didn't see why he was angry about it.  Maybe he was annoyed that I was following the common practice, of not worrying about the fact that the renormalization constants I encountered were infinite.  Torsten Gustafson\footnote{T. Gustafson, Nucl. Phys. A140, 1 (1970).} has said of K\"{a}ll\'{e}n that ``Like Pauli he often expressed his opinion in a provocative fashion --- especially to well-known physicists.''  I certainly was not a well-known physicist, but maybe K\"{a}ll\'{e}n was paying me a compliment by treating me like one.

I did not meet  K\"{a}ll\'{e}n again after this, and I never replied to his note.  I regret that very much,  because I think that if I had replied we could have understood each other, and been friends again.  Perhaps this talk can substitute for the reply to  K\"{a}ll\'{e}n that I should have made half a century ago.

\vspace{25pt}

I am grateful to C. Jarlskog and the  K\"{a}ll\'{e}n Lecture Committee for inviting me to Lund to give this talk, and to the Gunnar and Gunnel K\"{a}ll\'{e}n Memorial Fund of the Royal Physiographic Society for sponsoring it.  This material is based in part upon work supported by the National Science Foundation under Grant No. PHY-0455649 and with support from The Robert A. Welch Foundation, Grant No. F-0014.

\end{document}